\documentclass[english,15pt]{article}
\usepackage[T1]{fontenc}
\usepackage[latin1]{inputenc}
\usepackage{geometry}
\usepackage{float}
\usepackage{mathrsfs}
\usepackage{amsmath}
\usepackage{amssymb}
\usepackage{graphicx}
\usepackage{subfigure}

\makeatletter

\usepackage{algorithm,algpseudocode}

\usepackage{amsthm}

\usepackage{mathrsfs}

\usepackage{amsfonts}

\usepackage{epsfig}

\usepackage{bm}

\usepackage{mathrsfs}

\usepackage{enumerate}

\numberwithin{equation}{section}

\@ifundefined{definecolor}{\@ifundefined{definecolor}
 {\@ifundefined{definecolor}
 {\usepackage{color}}{}
}{}
}{}

\usepackage{subfig}\usepackage[all]{xy}

\newcounter{hypA}

\newcounter{hypB}

\newcounter{hypD}

\usepackage{babel}\date{}

\usepackage{babel}

\makeatother

\usepackage{babel}

\begin{document}

\begin{center}

{\Large \textbf{Bayesian Parameter Inference for Partially Observed SDEs driven by Fractional Brownian Motion}}

\vspace{0.5cm}

MOHAMED MAAMA$^{1,2}$, AJAY JASRA$^{1}$, \& HERNANDO OMBAO$^{2}$

{\footnotesize Applied Mathematics and Computational Science Program$^{1}$ \& Statistics Program$^{2}$, \\ Computer, Electrical and Mathematical Sciences and Engineering Division, \\ King Abdullah University of Science and Technology, Thuwal, 23955, KSA.} \\
{\footnotesize E-Mail:\,} \texttt{{\footnotesize maama.mohamed@gmail.com, ajay.jasra@kaust.edu.sa, hernando.ombao@kaust.edu.sa}} \\
\end{center}

\begin{abstract}
In this paper we consider Bayesian parameter inference for partially observed fractional Brownian motion (fBM) models. 
The approach we follow is to time-discretize the hidden process and then to design Markov chain Monte Carlo (MCMC) algorithms to sample from the posterior density on the parameters given data. We rely on a novel representation of the time discretization, which seeks to sample from an approximation of the posterior and then corrects via importance sampling; the approximation reduces the time (in terms of total observation time $T$) by $\mathcal{O}(T)$. This method is extended by using a multilevel MCMC method which can reduce the computational cost to achieve a given mean square error (MSE) versus using a single time discretization. Our methods are illustrated on simulated and real data.
\end{abstract}

\section{Introduction}

We consider a stochastic process on a given time interval $[0,T]$,  $T\in\mathbb{N}$,  defined on a probability space $(\Omega, \mathcal{F},\mathbb{P}_{\theta})$,  $\theta\in\Theta\subseteq\mathbb{R}^{d_{\theta}}$ is a collection of parameters of interest,  that follows the dynamics,    
\begin{equation}\label{eq:pofbm}
dX_t = a_{\theta}(X_t)dt + \sigma_{\theta}(X_t)dB_t^H
\end{equation}
where, for each $t\in[0,T]$, $X_t\in\mathbb{R}^d$, with $X_0$ given,   $a:\Theta\times\mathbb{R}^d\rightarrow\mathbb{R}^d$, $\sigma:\Theta\times\mathbb{R}^d\rightarrow\mathbb{R}^{d\times d}$ and $\{B_t^H\}_{t\in[0,T]}$ is a fractional Brownian motion in $d-$dimensions ($d-$one dimensional independent fractional Brownian motions) with Hurst parameter $H\in(0,1)$. This process will be observed through data that are recorded at discrete and regular time-instances.  This collection of models can be considered as a partially observed long-memory process and is of interest in a wide-class of applications such as in financial modeling; see for instance \cite{beskos_fbm,gander_long_mem}. This class of models are termed partially observed SDEs driven by fraction Brownian motion and we abbreviate as POfBM.

We consider the Bayesian inferential perspective with priors on the parameters and the objective is joint parameter and state inference. By state inference, we mean estimating the hidden process $\{X_t\}_{t\in[0,T]}$, typically at some discrete times and conditional upon the data.  Statistical inference for such models can be rather challenging. Other than the usual intractability of the posterior density w.r.t.~the unknown marginal likelihood, the finite dimensional density of \eqref{eq:pofbm}, over a time grid, is seldom available in closed form if it exists. In the scenario that a convergent time discretization of \eqref{eq:pofbm} is available, one is left with a non-Markovian state-space `type' model in high-frequency but discrete time. Even in the case that the hidden dynamics (i.e.~a process such as \eqref{eq:pofbm}) is Markovian, Bayesian inference still requires some ingenuity as we now discuss.

The case of Bayesian inference for hidden Markov models, is complicated even when the state-dynamics are in low dimensions, as one often has a target (posterior) space of the type
$\mathbb{R}^{dT}\times\Theta$. This high-dimensional space and the often extreme dependence between the hidden states and unknown parameter can make conventional Markov chain Monte Carlo algorithms ineffective. Several more advanced algorithms, by now, have become the gold standard including \cite{andrieu,delig_2018} and these are the methodologies (particle MCMC) that we will focus upon In the presence of time-discretization associated to the simulation of a probability, one method which can significantly enhance standard Monte Carlo methods is the multilevel Monte Carlo (MLMC) method of \cite{giles,hein} (see also \cite{giles1}). This is a Monte Carlo approximation method which works with a telescoping sum of expectations w.r.t.~the probability laws associated to increasingly more precise time discretizations. Given an appropriate coupling of the time-discretized probabilities which are `close', one can reduce, versus simulation from the most precise time-discretization, the computational effort to achieve a pre-specified mean square error. This method was originally constructed in the case that the probability laws of interest can be directly sampled, which, to extend to our model of interest, cannot be performed. However, we several approaches for non-i.i.d.~simulation have been made in \cite{jasra_bpe_sde,mlpf,mlpf1}; see \cite{ml_rev} for a review. The article \cite{jasra_bpe_sde} constructs a particle MCMC method which can approximate the multilevel identity (telescoping sum) for partially observed SDE models driven by Brownian motion and the main focus of this article is to extend this approach to our class of partially observed SDE models driven by fractional Brownian motion.

The extension of \cite{jasra_bpe_sde} to POfBM models requires some attention. For instance, particle MCMC is a major component of \cite{jasra_bpe_sde} and these algorithms are driven by the particle filter; as commented by \cite{beskos_fbm} the latter is not designed for long-memory processes; nonetheless and as dismissed in the afore-mentioned article, the quality of the particle filter on the so-called path space is not primarily an issue as it is the ability to produce single state-trajectories which represent the posterior which is important. As a result, we persist with the methodology of \cite{andrieu,delig_2018} in this article. Another aspect is a convergent time-discretization; as discussed in detail in \cite[Section 2.3]{beskos_fbm}, standard time-discretizations do not necessarily converge in some aspects; we use the Euler method adapted if needed and close to those used in \cite{bayer_rough_path_ml}. The main contribution of this article is establish how the approach of \cite{jasra_bpe_sde}  can be extended to the POfBM models.
A standard application, which works directly with the method of \cite{jasra_bpe_sde}  would yield a cost of
$\mathcal{O}(T^2\Delta_l^{-1}\log(\Delta_l^{-1}))$, where $\Delta_l=2^{-l}$ is the time discretization.
We show, using so-called \emph{pseudo increments} of fBM that this cost can be reduced to $\mathcal{O}(T\Delta_l^{-1}\log(\Delta_l^{-1}))$.
In addition, we conjecture based upon the theory in \cite{jasra_bpe_sde,bayer_rough_path_ml} when using the circulant-based simulation of the finite dimensional distribution of fBM
(see e.g.~\cite{gauss}), that to obtain an MSE of $\mathcal{O}(\epsilon^2)$ that the associated computational effort is $\mathcal{O}(\epsilon^{-3}|\log(\epsilon)|)$, for some $\epsilon\in(0,1)$ given. This is illustrated in simulations for simulated and real data. 
We note that using a single level MCMC method, our results indicate that for $\mathcal{O}(\epsilon^2)$ MSE that the cost  is $\mathcal{O}(\epsilon^{-4})$ (at least up-to log factors).
We mention that the approach of \cite{jasra_bpe_sde} has been extended by \cite{chada_ub} to remove discretization bias entirely, which our method does not, but it could be adapted using the afore-mentioned ideas.

This article is structured as follows. In Section \ref{sec:model} we describe the model in detail. In Section \ref{sec:comp_inf} our computational strategy, along with some theoretical discussion is discussed. Finally in Section \ref{sec:sim_res} we present our simulation results.

\section{Model}\label{sec:model}

\subsection{Posterior Density}

We assume a data model, with observations available at unit times $1,2,\dots,T$ and observed on $\mathbb{R}^{d_y}$.  The choice of unit times is purely for notational convenience
and could be achieved for any regularly and discretely (in-time) observed data by time rescaling.
We will assume that this model is defined for $t\in\{1,\dots,T\}$ such that for any $A\in\mathcal{B}(\mathbb{R}^{d_y})$
$$
\mathbb{P}_{\theta}(Y_t\in A|\{Y_k\}_{k\in\{1,\dots,T\}\setminus\{t\}},\{X_s\}_{s\in[0,T]}) = \int_A g_\theta(y|x_t)dy
$$
with $dy$ the $d_y-$Lebesgue measure and $g:\Theta\times\mathbb{R}^{d}\times\mathbb{R}^{d_y}\rightarrow\mathbb{R}^+$ is a probability density.  Writing the law of process of the finite skeleton $X_1,\dots,X_T$ as $p_{\theta}(dx_{1:T})$, we can write the measure of $(Y_{1:T},X_{1:T})$ as
$$
p_{\theta}(d(y_{1:T},x_{1:T})) = \left\{\prod_{t=1}^T g_{\theta}(y_t|x_t)dy_t\right\} p_{\theta}(dx_{1:T}).
$$
We are interested in the posterior
$$
\pi(d(\theta,x_{1:T})|y_{1:T}) \propto p_{\theta}(y_{1:T}|x_{1:T})p_{\theta}(dx_{1:T})p(\theta)d\theta
$$
where $p_{\theta}(y_{1:T}|x_{1:T})=\prod_{t=1}^T g_{\theta}(y_t|x_t)$.

In practice $p_{\theta}(dx_{1:T})$ may not admit a density (although we will assume that it does) w.r.t.~Lebesgue measure and even so, the density
is often unavailable analytically for most problems of practical interest.  However, one can induce a simple Euler approximation of \eqref{eq:pofbm}, by setting $\Delta_l=2^{-l}$ and for $k\in\{0,1,\dots,T\Delta_l^{-1}-1\}$
\begin{equation}\label{eq:pofbm_disc}
X_{(k+1)\Delta_l} = X_{k\Delta_l} + a_{\theta}(X_{k\Delta_l})\Delta_l + \sigma_{\theta}(X_{k\Delta_l})\{B_{(k+1)\Delta_l}^H-B_{k\Delta_l}^H\}.
\end{equation}
As noted in \cite[Section 2.3]{beskos_fbm}, one must be careful with such a time-discretization and, in 1-dimension (i.e.~$d=1$ in \eqref{eq:pofbm}) it is advisable to use a Lamperti-type transformation to stabilize the discretization. We will constrain our simulation results to such contexts, but note that several discretizations can be found which are convergent; see e.g.~\cite{bayer_rough_path_ml}. Thus, for clarity, from herein we assume that $d=1$, but this is for exposition purposes only and generalizations can be made.
Note that the fBM can be simulated exactly with a cost $\mathcal{O}(T\Delta_l^{-1}\log(T\Delta_l^{-1}))$.
This leads a joint probability density
$$
p_{\theta}^l(y_{1:T},x_{\Delta_l:T}) = \left\{\prod_{t=1}^T g_{\theta}(y_t|x_t)\right\} p_{\theta}^l(x_{\Delta_l:T})
$$
with $p_{\theta}^l$ the density induced by \eqref{eq:pofbm_disc} and $x_{\Delta_l:T}=(x_{\Delta_l},x_{2\Delta_l},\dots,x_T)^{\top}$. We will thus seek to work with the posterior
$$
\pi^l(\theta,x_{\Delta_l:T}|y_{1:T}) \propto p_{\theta}^l(y_{1:T},x_{\Delta_l:T})p(\theta).
$$

\subsection{Inferential Objectives and Couplings}

The objective of this article is ultimately to compute expectations w.r.t.~the posterior. That is, for $l\in\mathbb{N}$ fixed  and $\varphi:\Theta\times\mathbb{R}^{dT}\rightarrow\mathbb{R}$ that is $\pi^l-$integrable, to compute
$$
\mathbb{E}_{\pi^l}[\varphi(\theta,X_{1:T})] = \int_{\Theta\times\mathbb{R}^{dT}} \varphi(\theta,x_{1:T})\pi^l(\theta,x_{\Delta_l:T}|y_{1:T}) d(\theta,x_{\Delta_l:T})
$$
where we use the notation $x_{1:T}=(x_1,\dots,x_T)^{\top}$. In addition, we also seek to approximate the multilevel identity which for $L\in\mathbb{N}$ given reads
\begin{equation}\label{eq:ml_id}
\mathbb{E}_{\pi^L}[\varphi(\theta,X_{1:T})] = \mathbb{E}_{\pi^0}[\varphi(\theta,X_{1:T})] + \sum_{l=1}^L \left\{\mathbb{E}_{\pi^l}[\varphi(\theta,X_{1:T})]-\mathbb{E}_{\pi^{l-1}}[\varphi(\theta,X_{1:T})]\right\}.
\end{equation}
The computational advantages of Monte Carlo approximation of the R.H.S.~are normally only realized if one can sample from an `appropriate' coupling of $(\pi_l,\pi_{l-1})$; the term `appropriate' is by now well understood and we direct the reader to \cite{ml_rev} for the details.

In order to produce a method whose cost, for a given $\pi_l$, will scale as $\mathcal{O}(\Delta_l^{-1}\log(\Delta_l^{-1}))$ we will need to employ a reparameterization approach as was originally presented in \cite{beskos_fbm}. In order to do this, we will need some notation. First, for \eqref{eq:pofbm_disc} with $x_0$ and the fBM increments $B_{\Delta_l}^H,B_{2\Delta_l}^H-B_{\Delta_l}^H,\dots,B_{1}^H-B_{1-\Delta_l}^H$ given, we write
$$
X_1 = F_{\theta}^l(x_0,B_{\Delta_l:1}^H)
$$
as the iterated (and deterministic) mapping induced by $\Delta_l^{-1}$ applications of \eqref{eq:pofbm_disc}. We note that for any $t\in\{2,\dots,T\}$ with $x_{t-1}$ and skeleton $B_{t-1+\Delta_l:t}^H$ given we can write $X_t= F_{\theta}^l(x_{t-1},B_{t-1+\Delta_l:t}^H)$. Second, we note that the standard Davies \& Harte method of simulation of fBM increments
can be written in the following manner. For $Z_{1:2\Delta_l^{-1}}\sim\mathcal{N}_{2\Delta_l^{-1}}(0,I)$ (where $\mathcal{N}_{2\Delta_l^{-1}}(0,I)$ is the $2\Delta_l^{-1}-$dimensional
Gaussian distribution with 0 mean and identity covariance matrix) there exists a linear mapping $A_l$ such that we can write $B_{\Delta_l:1}^H=A_lZ_{2\Delta_l^{-1}}$; details are given in \cite{beskos_fbm,gauss}. In addition, for $Z_{1:2\Delta_l^{-1},1:T}\sim\mathcal{N}_{2\Delta_l^{-1}T}(0,I)$ one can produce fBM $B_{\Delta_l:T}^H=A_{l,T}Z_{1:2\Delta_l^{-1},1:T}$.
At this stage, one could write
$$
\tilde{\pi}^l(\theta,z_{1:2\Delta_l^{-1},1:T}) \propto \left\{\prod_{t=1}^T g_{\theta}(y_t|F^{l,t}(x_0,G_T^{l,t}(z_{1:2\Delta_l^{-1},1:t}))\right\} p(z_{1:2\Delta_l^{-1},1:T}) p(\theta)
$$
where independently for each $t\in\{1,\dots,T\}$, $Z_{1:2\Delta_l^{-1},t}\sim \mathcal{N}_{2\Delta_l^{-1}}(0,I)$, $F^{l,t}$ represents the $t-$fold application of $F^l$
and $G_T^{l,t}(z_{1:2\Delta_l^{-1},1:t})=B_{\Delta_l:t}^H$ induced by the mapping $A_{l,T}Z_{1:2\Delta_l^{-1},1:T}$.
Given this representation, one would have the formula
\begin{equation}\label{eq:level0_id}
\mathbb{E}_{\pi^l}[\varphi(\theta,X_{1:T})] = \int_{\Theta\times\mathbb{R}^{2T\Delta_l^{-1}}} \tilde{\varphi}^l_l(\theta,z_{1:2\Delta_l^{-1},1:T})\tilde{\pi}^l(\theta,z_{1:2\Delta_l^{-1},1:T})d(\theta,z_{1:2\Delta_l^{-1},1:T})
\end{equation}
where $\tilde{\varphi}^l_l(\theta,z_{1:2\Delta_l^{-1},1:T})=\varphi(\theta,F^{l,t}(x_0,G^{l,1}(z_{1:2\Delta_l^{-1},1})),\dots,F^{l,T}(x_0,G^{l,T}(z_{1:2\Delta_l^{-1},1:T})))$. However,
we shall need to go further in-order to construct achieve our objectives.

Now, if one has $B_{\Delta_l}^H,B_{2\Delta_l}^H-B_{\Delta_l}^H,\dots,B_{T}^H-B_{T-\Delta_l}^H$ given, we can produce $B_{\Delta_{l-1}}^H,B_{2\Delta_{l-1}}^H-B_{\Delta_{l-1}}^H,\dots,B_{T}^H-B_{1-\Delta_{T-1}}^H$ simply by summing the appropriate increments of $B_{\Delta_l}^H,B_{2\Delta_l}^H-B_{\Delta_l}^H,\dots,B_{T}^H-B_{R-\Delta_l}^H$.
This is essentially related to the synchronous coupling that one uses for SDEs and is the one which we shall apply, in order to approximate the multilevel identity. We will
use the notation $H_T^{l-1,t}(z_{1:2\Delta_l^{-1},1:t})$ to represent first computing all the fBM increments $B_{\Delta_l}^H,B_{2\Delta_l}^H-B_{\Delta_l}^H,\dots,B_{t}^H-B_{t-\Delta_l}^H$
and then summing to get those at level $l-1$.

We also seek to employ the approach that is used in \cite{jasra_bpe_sde} so that end, we shall define a probability density which provides a different, but equal, representation of
$\mathbb{E}_{\pi_l}[\varphi(\theta,X_{1:T})]-\mathbb{E}_{\pi_{l-1}}[\varphi(\theta,X_{1:T})]$.
We will consider simulation from the following probability density:
\begin{eqnarray*}
\check{\pi}^{l}(\theta,z_{1:2\Delta_l^{-1},1:T}) & \propto &\left\{\prod_{t=1}^T \max\{g_{\theta}(y_t|F^{l,t}(x_0,G^{l,t}(z_{1:2\Delta_l^{-1},1:t}))),
g_{\theta}(y_t|F^{l-1,t}(x_0,H^{l-1,t}(z_{1:2\Delta_l^{-1},1:t})))
\}\right\} \times \\& & p(z_{1:2\Delta_l^{-1},1:T}) p(\theta)
\end{eqnarray*}
where $G^{l,t}(z_{1:2\Delta_l^{-1},1:t})=(A_lz_{1:2\Delta_l^{-1},1},\dots,A_lz_{1:2\Delta_l^{-1},t})^{\top}$, \emph{which is not} a discrete skeleton of fBM on $[0,t]$ and $H^{l-1,t}(z_{1:2\Delta_l^{-1},1:t})$
are the related summed `pseudo' increments.
Now, defining 
\begin{eqnarray*}
J_l^l(\theta,z_{1:2\Delta_l^{-1},1:T}) & = & \frac{\prod_{t=1}^T g_{\theta}(y_t|F^{l,t}(x_0,G_T^{l,t}(z_{1:2\Delta_l^{-1},1:t}))}{\prod_{t=1}^T \max\{g_{\theta}(y_t|F^{l,t}(x_0,G^{l,t}(z_{1:2\Delta_l^{-1},1:t})),
g_{\theta}(y_t|F^{l-1,t}(x_0,H^{l-1,t}(z_{1:2\Delta_l^{-1},1:t}))\}} \\
J_l^{l-1}(\theta,z_{1:2\Delta_l^{-1},1:T}) & = & \frac{\prod_{t=1}^T g_{\theta}(y_t|F^{l-1,t}(x_0,H_T^{l-1,t}(z_{1:2\Delta_l^{-1},1:t}))}{\prod_{t=1}^T \max\{g_{\theta}(y_t|F^{l,t}(x_0,G^{l,t}(z_{1:2\Delta_l^{-1},1:t})),
g_{\theta}(y_t|F^{l-1,t}(x_0,H^{l-1,t}(z_{1:2\Delta_l^{-1},1:t}))\}}
\end{eqnarray*}
it is simple to show that
$$
\mathbb{E}_{\pi^l}[\varphi(\theta,X_{1:T})]-\mathbb{E}_{\pi^{l-1}}[\varphi(\theta,X_{1:T})] = 
$$
\begin{equation}\label{eq:main_id}
\frac{\mathbb{E}_{\check{\pi}^l}[\tilde{\varphi}^l_l(\theta,Z_{1:2\Delta_l^{-1},1:T})J_l^l(\theta,Z_{1:2\Delta_l^{-1},1:T})]}{\mathbb{E}_{\check{\pi}^l}[J_l^l(\theta,Z_{1:2\Delta_l^{-1},1:T})]} - \frac{
\mathbb{E}_{\check{\pi}^l}[\tilde{\varphi}^{l-1}_l(\theta,Z_{1:2\Delta_l^{-1},1:T})J_l^{l-1}(\theta,Z_{1:2\Delta_l^{-1},1:T})]
}{\mathbb{E}_{\check{\pi}^l}[J_l^{l-1}(\theta,Z_{1:2\Delta_l^{-1},1:T})]}.
\end{equation}
where $\tilde{\varphi}^{l-1}_l(\theta,Z_{1:2\Delta_l^{-1},1:T})=\varphi(\theta,F^{l-1,1}(x_0,H^{l-1,t}(z_{1:2\Delta_l^{-1},1})),\dots,F^{l-1,T}(x_0,H^{l-1,T}(z_{1:2\Delta_l^{-1},1:T})))$.

Given the identities \eqref{eq:level0_id} and \eqref{eq:main_id} our objective is now to find sampling methods for $\tilde{\pi}^0$ and $\check{\pi}_l$, for $l\in\{1,\dots,L\}$. 
The precise reason for using the afore-mentioned identities is explained in details in \cite{jasra_bpe_sde}. The approach uses pseudo-increments of fBM to cut the cost of the resulting algorithm as a function of $T$. If one used exact increments of fBM the cost of simulation (per-iteration of an MCMC) would be at least $\mathcal{O}(T^2\Delta_l^{-1}\log(\Delta_l^{-1}))$, however, using these pseudo-increments of fBM  the cost is 
now $\mathcal{O}(T\Delta_l^{-1}\log(\Delta_l^{-1}))$. The inevitable price to pay are modified importance weights 
e.g.~$J_l^l(\theta,z_{1:2\Delta_l^{-1},1:T})$ which may be less stable; however, we find in simulation that this potential degradation does not overly effect our results.

\section{Computational Inference}\label{sec:comp_inf}

\subsection{Particle Filters}

In order to sample from  $\tilde{\pi}^l$ and $\check{\pi}^l$ we shall need two particle filter algorithms which are presented in Algorithms \ref{alg:pf1} and \ref{alg:pf2}. These two algorithms are needed for the first level and then for simulation across levels.

\begin{algorithm}[h]
\begin{enumerate}
\item{Input: level $l$, parameter $\theta$ and number of particles $N$.}
\item{Initialize. For $i\in\{1,\dots,N\}$ sample $Z_{1:2\Delta_l^{-1},1}^i\sim\mathcal{N}_{2\Delta_l^{-1}}(0,I)$. Compute the un-normalized weight
$\tilde{w}_1^i = g_{\theta}(y_1|F^{l,1}(x_0,G^{l,1}(z_{1:2\Delta_l^{-1},1}^i))$ and the normalizing constant estimator $\tilde{C}_{\theta,1}^{N,l}=\tfrac{1}{N}\sum_{j=1}^N \tilde{w}_1^i$. Set $t=1$.
}
\item{Iterate. Resample $Z_{1:2\Delta_l^{-1},1:t}^{1:N}$ using the normalized weights $\left(\tilde{w}_t^1/\sum_{j=1}^N\tilde{w}_t^j,\dots,\tilde{w}_t^N/\sum_{j=1}^N\tilde{w}_t^j\right)$ calling the
resulting samples $Z_{1:2\Delta_l^{-1},1:t}^{1:N}$. For $i\in\{1,\dots,N\}$ sample $Z_{1:2\Delta_l^{-1},t+1}^i\sim\mathcal{N}_{2\Delta_l^{-1}}(0,I)$.
Compute the un-normalized weight $\tilde{w}_{t+1}^i = g_{\theta}(y_{t+1}|F^{l,t+1}(x_0,G^{l,t+1}(z_{1:2\Delta_l^{-1},1:t+1}^i))$
and the normalizing constant estimator $\tilde{C}_{\theta,t+1}^{N,l}=\tilde{C}_{\theta,t}^{N,l}\tfrac{1}{N}\sum_{j=1}^N \tilde{w}_{t+1}^i$.
Set $t=t+1$ and $t=T+1$ go to 4., otherwise go to the start of step 2..}
\item{Sample a single trajectory $z_{1:2\Delta_l^{-1},1:T}$ from $z_{1:2\Delta_l^{-1},1:T}^{1:N}$ using the normalized weights
$\left(\tilde{w}_T^1/\sum_{j=1}^N\tilde{w}_T^j,\dots,\tilde{w}_T^N/\sum_{j=1}^N\tilde{w}_T^j\right)$. Go to 5..}
\item{Output: normalizing constant estimate $\tilde{C}_{\theta,T}^{N,l}$ and $z_{1:2\Delta_l^{-1},1:T}$.}
\end{enumerate}
\caption{Particle Filter associated to  $\tilde{\pi}^l$.}
\label{alg:pf1}
\end{algorithm}

\begin{algorithm}[h]
\begin{enumerate}
\item{Initialize. level $l$, parameter $\theta$ and number of particles $N$.}
\item{Initialize. For $i\in\{1,\dots,N\}$ sample $Z_{1:2\Delta_l^{-1},1}^i\sim\mathcal{N}_{2\Delta_l^{-1}}(0,I)$. Compute the un-normalized weight
$$
\check{w}_1^i = \max\{g_{\theta}(y_t|F^{l,1}(x_0,G^{l,t}(z_{1:2\Delta_l^{-1},1}^i))),
g_{\theta}(y_t|F^{l-1,1}(x_0,H^{l-1,1}(z_{1:2\Delta_l^{-1},1}^i)))
\}
$$
and the normalizing constant estimator $\check{C}_{\theta,1}^{N,l}=\tfrac{1}{N}\sum_{j=1}^N \check{w}_1^i$. Set $t=1$.
}
\item{Iterate. Resample $Z_{1:2\Delta_l^{-1},1:t}^{1:N}$ using the normalized weights $\left(\check{w}_t^1/\sum_{j=1}^N\check{w}_t^j,\dots,\check{w}_t^N/\sum_{j=1}^N\check{w}_t^j\right)$ calling the
resulting samples $Z_{1:2\Delta_l^{-1},1:t}^{1:N}$. For $i\in\{1,\dots,N\}$ sample $Z_{1:2\Delta_l^{-1},t+1}^i\sim\mathcal{N}_{2\Delta_l^{-1}}(0,I)$.
Compute the un-normalized weight 
$$
\check{w}_{t+1}^i = \max\{g_{\theta}(y_{t+1}|F^{l,t+1}(x_0,G^{l,t}(z_{1:2\Delta_l^{-1},1:t+1}^i))),
g_{\theta}(y_{t+1}|F^{l-1,t}(x_0,H^{l-1,t}(z_{1:2\Delta_l^{-1},1:t}^i)))
\}
$$
and the normalizing constant estimator $\check{C}_{\theta,t+1}^{N,l}=\check{C}_{\theta,t}^{N,l}\tfrac{1}{N}\sum_{j=1}^N \check{w}_{t+1}^i$.
Set $t=t+1$ and $t=T+1$ go to 4., otherwise go to the start of step 2..}
\item{Sample a single trajectory $z_{1:2\Delta_l^{-1},1:T}$ from $z_{1:2\Delta_l^{-1},1:T}^{1:N}$ using the normalized weights 
$\left(\check{w}_T^1/\sum_{j=1}^N\check{w}_T^j,\dots,\check{w}_T^N/\sum_{j=1}^N\check{w}_T^j\right)$. Go to 5..}
\item{Output: normalizing constant estimate $\check{C}_{\theta,T}^{N,l}$ and $z_{1:2\Delta_l^{-1},1:T}$.}
\end{enumerate}
\caption{Particle Filter associated to  $\check{\pi}^l$.}
\label{alg:pf2}
\end{algorithm}

\subsection{Particle MCMC}

We now present our methods to sample from  $\tilde{\pi}^l$ and $\check{\pi}_l$  in Algorithms \ref{alg:pmcmc1} and \ref{alg:pmcmc2}.

\begin{algorithm}[h]
\begin{enumerate}
\item{Input: level $l$, number of particles $N$ and number of iterations $M$.}
\item{Initialization: Sample $\theta^l(0)$ from the prior and run Algorithm \ref{alg:pf1} with level $l$, parameter $\theta^l(0)$ and $N$ particles. Denote the returned 
normalizing constant estimate $\tilde{C}_{\theta^l(0),T}^{N,l}$ and trajectory $z_{1:2\Delta_l^{-1},1:T}^l(0)$. Set $t=1$.}
\item{Iteration: Generate $\theta'$ from a proposal $q(\cdot|\theta^l(t-1))$ and run Algorithm \ref{alg:pf1} with level $l$, parameter $\theta'$ and $N$ particles
denoting the returned 
normalizing constant estimate $\tilde{C}_{\theta',T}^{N,l}$ and trajectory $z_{1:2\Delta_l^{-1},1:T}'$. Set $(\theta^l(t),z_{1:2\Delta_l^{-1},1:T}^l(t),\tilde{C}_{\theta^l(t),T})=(\theta',z_{1:2\Delta_l^{-1},1:T}',\tilde{C}_{\theta',T}^{N,l})$
with probability
$$
\min\left\{1,\frac{\tilde{C}_{\theta',T}^{N,l}p(\theta')q(\theta^l(t-1)|\theta')}{\tilde{C}_{\theta^l(t-1),T}^{N,l}p(\theta^l(t-1))q(\theta^l(t-1)|\theta')}\right\}
$$
otherwise set $(\theta^l(t),z_{1:2\Delta_l^{-1},1:T}^l(t),\tilde{C}_{\theta^l(t),T})=(\theta^l(t-1),z_{1:2\Delta_l^{-1},1:T}^l(t-1),\tilde{C}_{\theta^l(t-1),T})$. Set $t=t+1$ and if $t=M+1$ go to 4., otherwise go to the start of step 3..}
\item{Output: $(\theta^l(0:M),z_{1:2\Delta_l^{-1},1:T}^l(0:M))$.}
\end{enumerate}
\caption{Particle MCMC associated to  $\tilde{\pi}^l$.}
\label{alg:pmcmc1}
\end{algorithm}

\begin{algorithm}[h]
\begin{enumerate}
\item{Input: level $l$, number of particles $N$ and number of iterations $M$.}
\item{Initialization: Sample $\theta^l(0)$ from the prior and run Algorithm \ref{alg:pf2} with level $l$, parameter $\theta^l(0)$ and $N$ particles. Denote the returned 
normalizing constant estimate $\check{C}_{\theta^l(0),T}^{N,l}$ and trajectory $z_{1:2\Delta_l^{-1},1:T}^l(0)$. Set $t=1$.}
\item{Iteration: Generate $\theta'$ from a proposal $q(\cdot|\theta^l(t-1))$ and run Algorithm \ref{alg:pf2} with level $l$, parameter $\theta'$ and $N$ particles
denoting the returned 
normalizing constant estimate $\check{C}_{\theta',T}^{N,l}$ and trajectory $z_{1:2\Delta_l^{-1},1:T}'$. Set $(\theta^l(t),z_{1:2\Delta_l^{-1},1:T}^l(t),\check{C}_{\theta^l(t),T})=(\theta',z_{1:2\Delta_l^{-1},1:T}',\check{C}_{\theta',T}^{N,l})$
with probability
$$
\min\left\{1,\frac{\check{C}_{\theta',T}^{N,l}p(\theta')q(\theta^l(t-1)|\theta')}{\check{C}_{\theta^l(t-1),T}^{N,l}p(\theta^l(t-1))q(\theta^l(t-1)|\theta')}\right\}
$$
otherwise set $(\theta^l(t),z_{1:2\Delta_l^{-1},1:T}^l(t),\check{C}_{\theta^l(t),T})=(\theta^l(t-1),z_{1:2\Delta_l^{-1},1:T}^l(t-1),\check{C}_{\theta^l(t-1),T})$. Set $t=t+1$ and if $t=M+1$ go to 4., otherwise go to the start of step 3..}
\item{Output: $(\theta^l(0:M),z_{1:2\Delta_l^{-1},1:T}^l(0:M))$.}
\end{enumerate}
\caption{Particle MCMC associated to  $\check{\pi}^l$.}
\label{alg:pmcmc2}
\end{algorithm}

\subsection{Multilevel Method}

The approach that we will use is as below.
\begin{itemize}
\item{At level 0 run Algorithm \ref{alg:pmcmc1} with level 0, $N_0$ particles and $M_0$ iterations.}
\item{For each level $l\in\{1,\dots,L\}$, independently of each other and level 0, run Algorithm \ref{alg:pmcmc2} with level $l$, $N_l$ particles and $M_l$ iterations.}
\end{itemize}

Define
$$
J_0(\theta,z_{1:2\Delta_0^{-1},1:T}) = \prod_{t=1}^T\frac{g_{\theta}(y_t|F^{0,t}(x_0,G_T^{l,t}(z_{1:2\Delta_0^{-1},1:T})))}{g_{\theta}(y_t|F^{0,t}(x_0,G^{l,t}(z_{1:2\Delta_0^{-1},1:T})))}.
$$
Then, to approximate the multilevel identity \eqref{eq:ml_id} we have the estimator
\begin{eqnarray*}
\widehat{\mathbb{E}_{\pi^L}[\varphi(\theta,X_{1:T})]} & = & \frac{\tfrac{1}{M_0}\sum_{i=1}^{M_0} \tilde{\varphi}^0_0(\theta^0(i),z_{1:2\Delta_0^{-1},1:T}^0(i))J_0(\theta^0(i),z_{1:2\Delta_0^{-1},1:T}^0(i))}{\tfrac{1}{M_0}\sum_{i=1}^{M_0} J_0(\theta^0(i),z_{1:2\Delta_0^{-1},1:T}^0(i))} + \\
& & 
\sum_{l=1}^L\Bigg\{\frac{\tfrac{1}{M_l}\sum_{i=1}^{M_l}\tilde{\varphi}^l_l(\theta^l(i),z_{1:2\Delta_l^{-1},1:T}^l(i))J_l^l(\theta^l(i),z_{1:2\Delta_l^{-1},1:T}^l(i))}
{\tfrac{1}{M_l}\sum_{i=1}^{M_l}J_l^l(\theta^l(i),z_{1:2\Delta_l^{-1},1:T}^l(i))} - \\ & &
\frac{\tfrac{1}{M_l}\sum_{i=1}^{M_l}\tilde{\varphi}^{l-1}_l(\theta^l(i),z_{1:2\Delta_l^{-1},1:T}^l(i))J_l^{l-1}(\theta^l(i),z_{1:2\Delta_l^{-1},1:T}^l(i))}
{\tfrac{1}{M_l}\sum_{i=1}^{M_l}J_l^{l-1}(\theta^l(i),z_{1:2\Delta_l^{-1},1:T}^l(i))}
\Bigg\}.
\end{eqnarray*}
To shorten the subsequent notations we will set for $l\in\{1,\dots,L\}$
\begin{eqnarray*}
\widehat{\pi}^{M_l,l}(\varphi) & = & \frac{\tfrac{1}{M_l}\sum_{i=1}^{M_l}\tilde{\varphi}^l_l(\theta^l(i),z_{1:2\Delta_l^{-1},1:T}^l(i))J_l^l(\theta^l(i),z_{1:2\Delta_l^{-1},1:T}^l(i))}
{\tfrac{1}{M_l}\sum_{i=1}^{M_l}J_l^l(\theta^l(i),z_{1:2\Delta_l^{-1},1:T}^l(i))} \\
\widehat{\pi}^{M_l,l-1}(\varphi) & = & \frac{\tfrac{1}{M_l}\sum_{i=1}^{M_l}\tilde{\varphi}^{l-1}_l(\theta^l(i),z_{1:2\Delta_l^{-1},1:T}^l(i))J_l^{l-1}(\theta^l(i),z_{1:2\Delta_l^{-1},1:T}^l(i))}
{\tfrac{1}{M_l}\sum_{i=1}^{M_l}J_l^{l-1}(\theta^l(i),z_{1:2\Delta_l^{-1},1:T}^l(i))}.
\end{eqnarray*}

\subsection{Theory}\label{sec:theory}

In this Section we discuss the choice of $L$ and $M_l$, $l\in\{0,1,\dots,L\}$. The general way in which this would be done is to obtain bounds on
$$
|\mathbb{E}_{\pi}[\varphi(\theta,X_{1:T})] - \mathbb{E}_{\pi}[\varphi(\theta,X_{1:T})]|
$$
where $\pi$ is the joint posterior of the states and parameter from the exact model, i.e.~one without time discretization.
In the case of SDEs this is $\mathcal{O}(\Delta_L^{\alpha})$, $\alpha=1$; our simulations will indicate that $\alpha=1/2$, but we do not know this to be mathematically correct. The next result one would require would be a bound on
$$
\mathbb{E}\left[\left(
\widehat{\pi}^{M_l,l}(\varphi)-\widehat{\pi}^{M_l,l-1}(\varphi) - 
\{\mathbb{E}_{\pi^l}[\varphi(\theta,X_{1:T})] - \mathbb{E}_{\pi^{l-1}}[\varphi(\theta,X_{1:T})]\}
\right)^2\right].
$$
In the case of \cite{jasra_bpe_sde} the authors show that for SDEs driven by Brownian motion, these terms are
$$
\mathcal{O}\left(\frac{\Delta_l^{\beta}}{M_l}\right)
$$
where $\beta$ is the strong error rate associated to the time discretization. Although we do not have precise results on that, based upon the analysis in \cite{bayer_rough_path_ml} it seems that the above rate would be reasonable.
The simulations will indicate that $\beta=1/2$.
Given the afore-mentioned conjectures, \cite[Theorem 20]{bayer_rough_path_ml} indicate that to obtain a MSE of $\mathcal{O}(\epsilon^2)$, for $\epsilon\in(0,1)$ given, one can choose $L$ and $M_l$, $l\in\{0,1,\dots,L\}$ as \cite[eq.~(16)-(17)]{bayer_rough_path_ml}  and one would expect the cost is $\mathcal{O}(\epsilon^{-3}|\log(\epsilon)|)$.
In this case ($\alpha>\beta/2$ in \cite{bayer_rough_path_ml}, we assume $\beta=1/2$), \cite{bayer_rough_path_ml} show an $\mathcal{O}(\epsilon^{-1})$ speed up versus a single level, which according to our simulations would have a cost of $\mathcal{O}(\epsilon^{-4})$ for a MSE of $\mathcal{O}(\epsilon^2)$. 

\section{Simulation Results}\label{sec:sim_res}

\subsection{Model Settings}

We now consider two numerical examples associate to the OU type model driven by fBM:
$$
dX_t  = -\theta X_tdt + \sigma dB^H_t, \quad X_0=x_0
$$
with $d=d_y=1$.
In both examples to be considered, the observations are such that for $k\in\{1,\dots,T\}$, $Y_k|x_k,\theta\sim\mathcal{N}(x_k,\tau^2)$ the one-dimensional Gaussian with mean $x_k$ and variance $\tau^2$ and we set $\tau^2=0.2$, $H=0.4$.
The parameters to be estimated are then $\theta$ and $\sigma$ and these are both assigned independent Gamma priors.

In the simulated data case, 100 observations generated from the model under discretization, we assign priors that 
are $\mathcal{G}a(1,0.5)$ (Gamma distribution of shape 0.5 and scale 1) for $\sigma$ and $\mathcal{G}a(1,1)$ for $\theta$. In the case of real data daily from the S\&P $500$, which are the log-returns from January 4, 2021 to December 30, 2021, the
priors were $\mathcal{G}a(1/1000,1/1000)$ for both parameters.

\subsection{Simulation Settings}

In our multilevel method, we only allow $l\in\{3,4,\dots,7\}$ and $L$ and $M_l$ are chosen as in \cite{bayer_rough_path_ml} (see Section \ref{sec:theory}). The number of samples used in the particle filter for the particle MCMC algorithm is taken as $\mathcal{O}(T)$, which should control the variance of the acceptance probability.

\subsection{Numerical Results}

\begin{figure}[h]
\centering
\subfigure[Estimated States]{\includegraphics[width=12cm,height=5.5cm]{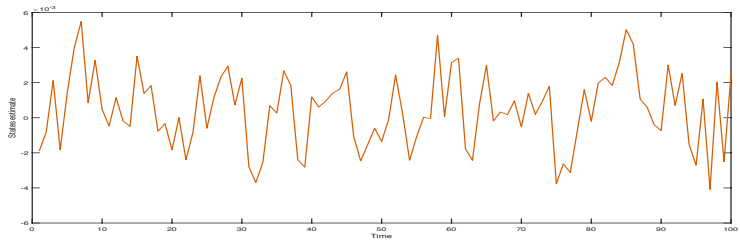}}
\subfigure[$\theta$]{\includegraphics[width=6.5cm,height=6.5cm]{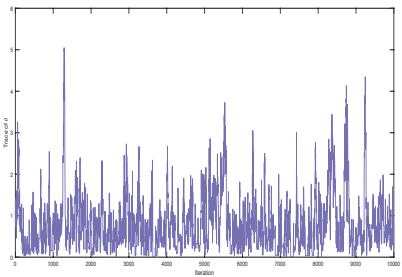}}
\subfigure[$\sigma$]{\includegraphics[width=6.5cm,height=6.5cm]{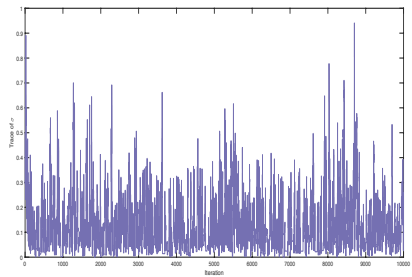}}
 \caption{Single Level Particle MCMC Results for Simulated Data. The simulations were at level 7.}
    \label{fig:OU}
\end{figure}

We begin by considering the simulated data.
In Figure \ref{fig:OU} we can observe some output from the single level MCMC algorithm run at level 7. We see the state-estimates and the trace plots from the chain. These indicate rather good mixing for this example, although we note of course that the samples here are not corrected by importance sampling.

\begin{figure}[h]
\centering
\subfigure[$\theta$]{\includegraphics[width=6.5cm,height=6.5cm]{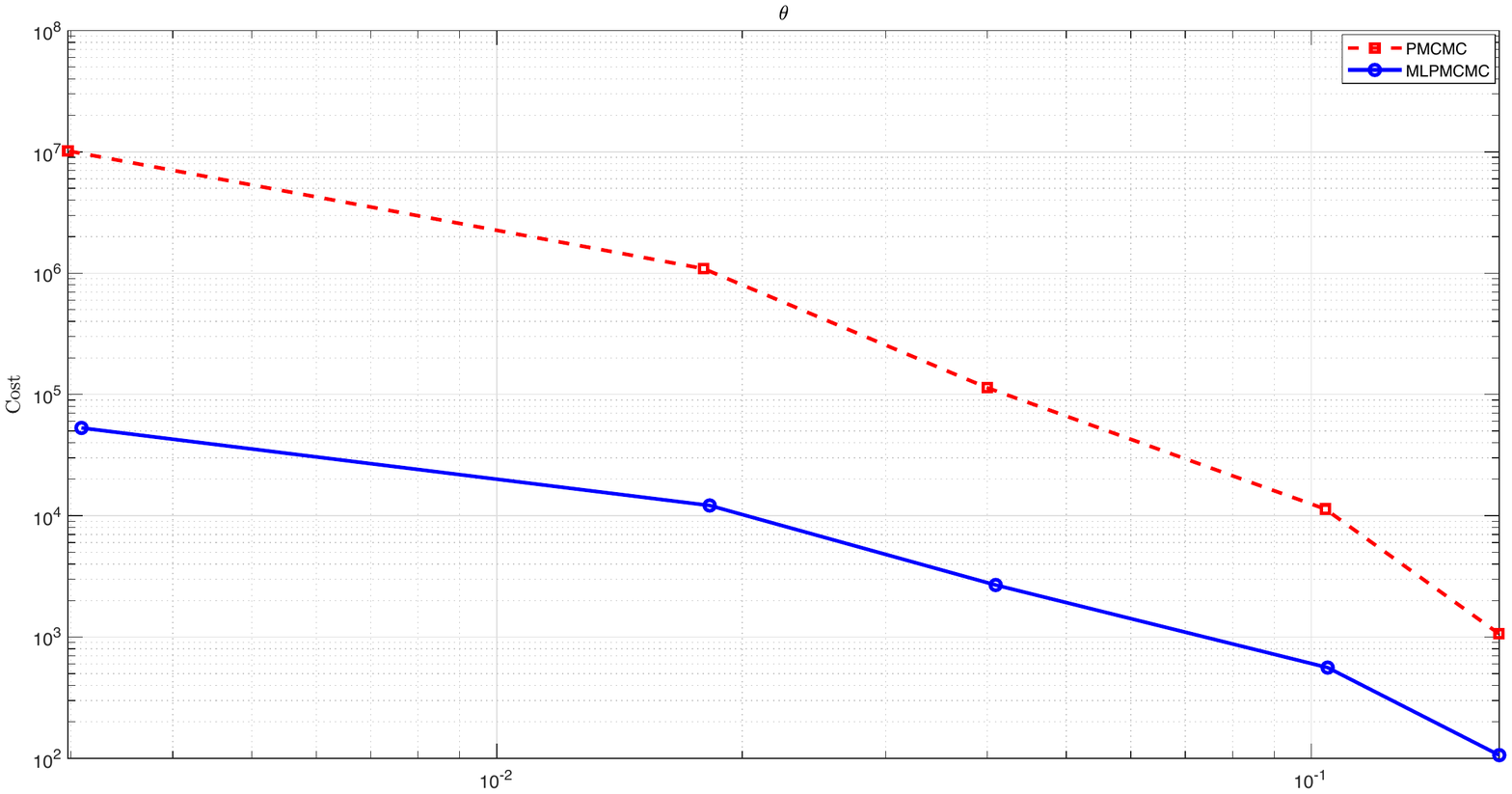}}
\subfigure[$\sigma$]{\includegraphics[width=6.5cm,height=6.5cm]{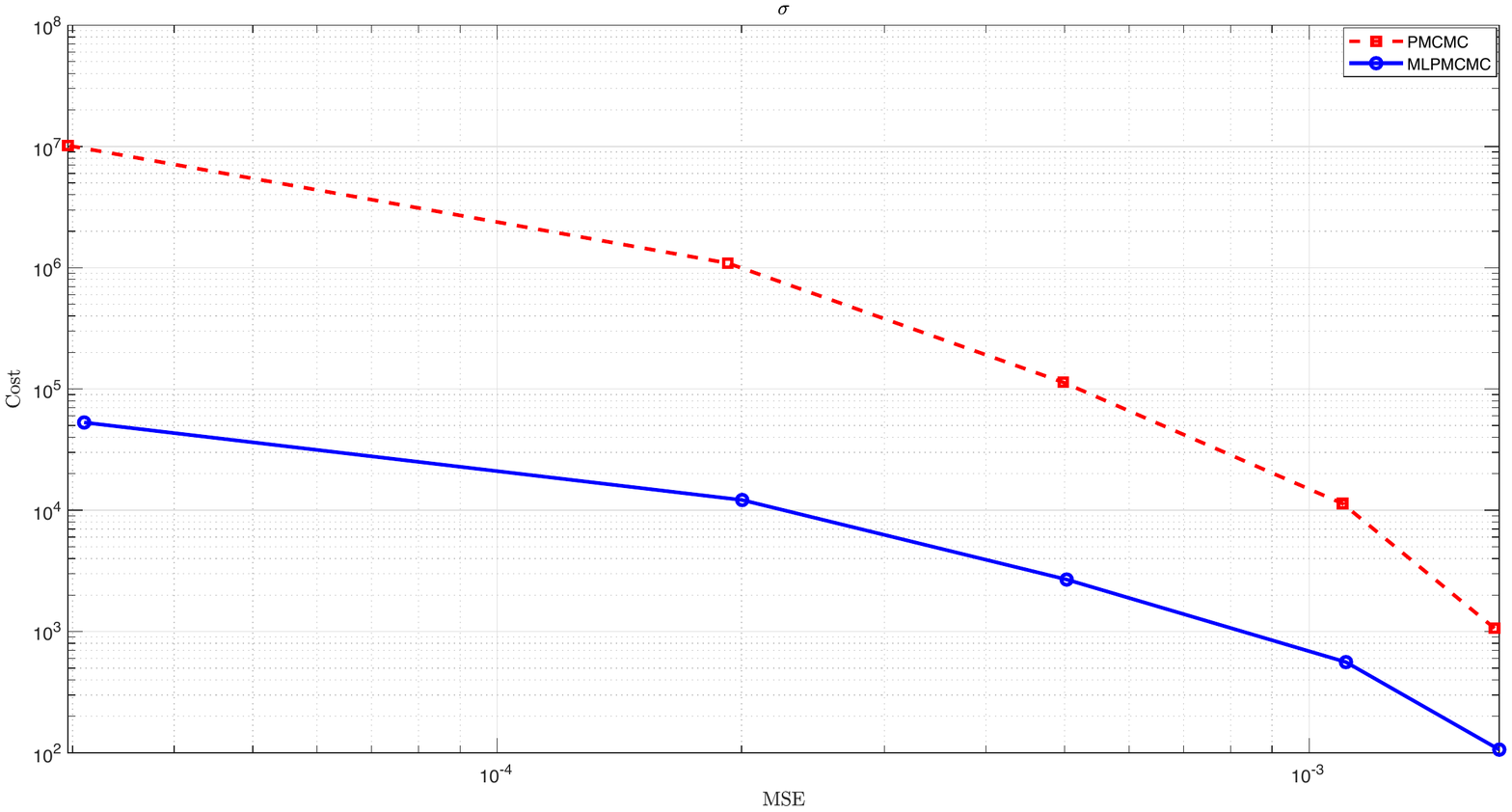}}
 \caption{Cost versus MSE Plots for Simulated Data. }
      \label{fig:msecost1}
\end{figure}

In Figure \ref{fig:msecost1} we can see the cost-MSE plots (based upon 50 repeats). They clearly indicate that the multilevel MCMC method has a lower cost to achieve a given MSE. In Table \ref{tab:res} 
we estimate the rates, that is, log cost against log MSE based upon Figure \ref{fig:msecost1}. This suggests that a single level method has a cost of $\mathcal{O}(\epsilon^{-4})$ to achieve an MSE of $\mathcal{O}(\epsilon^2)$ at least up-to log-factors. If one inspects Section \ref{sec:theory} this suggests that $\alpha=1/2$ and using \cite[Theorem 20]{bayer_rough_path_ml} along with Table \ref{tab:res}  that $\beta=1/2$. This is because Table  \ref{tab:res} says that
the cost is $\mathcal{O}(\epsilon^{-3})$ to achieve an MSE of $\mathcal{O}(\epsilon^2)$; again up-to logarithmic factors. 
 \cite[Theorem 20]{bayer_rough_path_ml} states (although does not directly apply to the problem in this article) then that the cost is as $\mathcal{O}(\epsilon^{-3}|\log(\epsilon)|)$.

\begin{table}[h]
\begin{center}
\begin{tabular}{ |c|c|c|c| } 
\hline
Data  & Parameter & PMCMC & MLPMCMC \\
\hline
Simulated & $\theta$ & -2.205 & -1.501 \\
 & $\sigma$ & -2.164 & -1.475 \\
 \hline
Real & $\theta$ & -2.191 & -1.492 \\
 & $\sigma$ & -2.153 & -1.473 \\
 \hline
\end{tabular}
\caption{Estimated log cost versus log of the MSE based upon the results from \ref{fig:msecost1} and \ref{fig:msecost2}.}\label{tab:res}
\end{center}
\end{table}

\begin{figure}[h]
\centering
\subfigure[Estimated States]{\includegraphics[width=12cm,height=5.5cm]{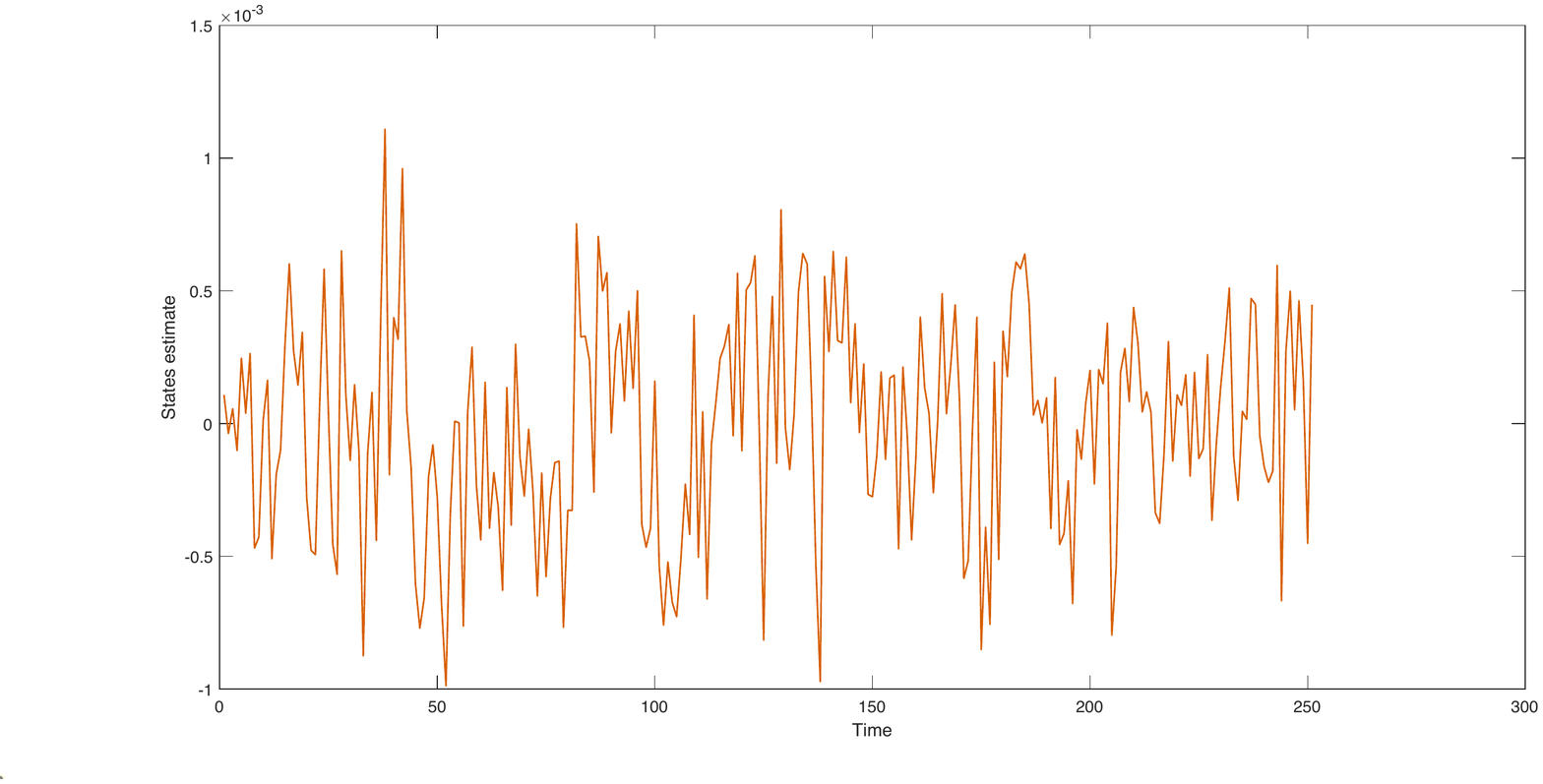}}
\subfigure[$\theta$]{\includegraphics[width=6.5cm,height=6.5cm]{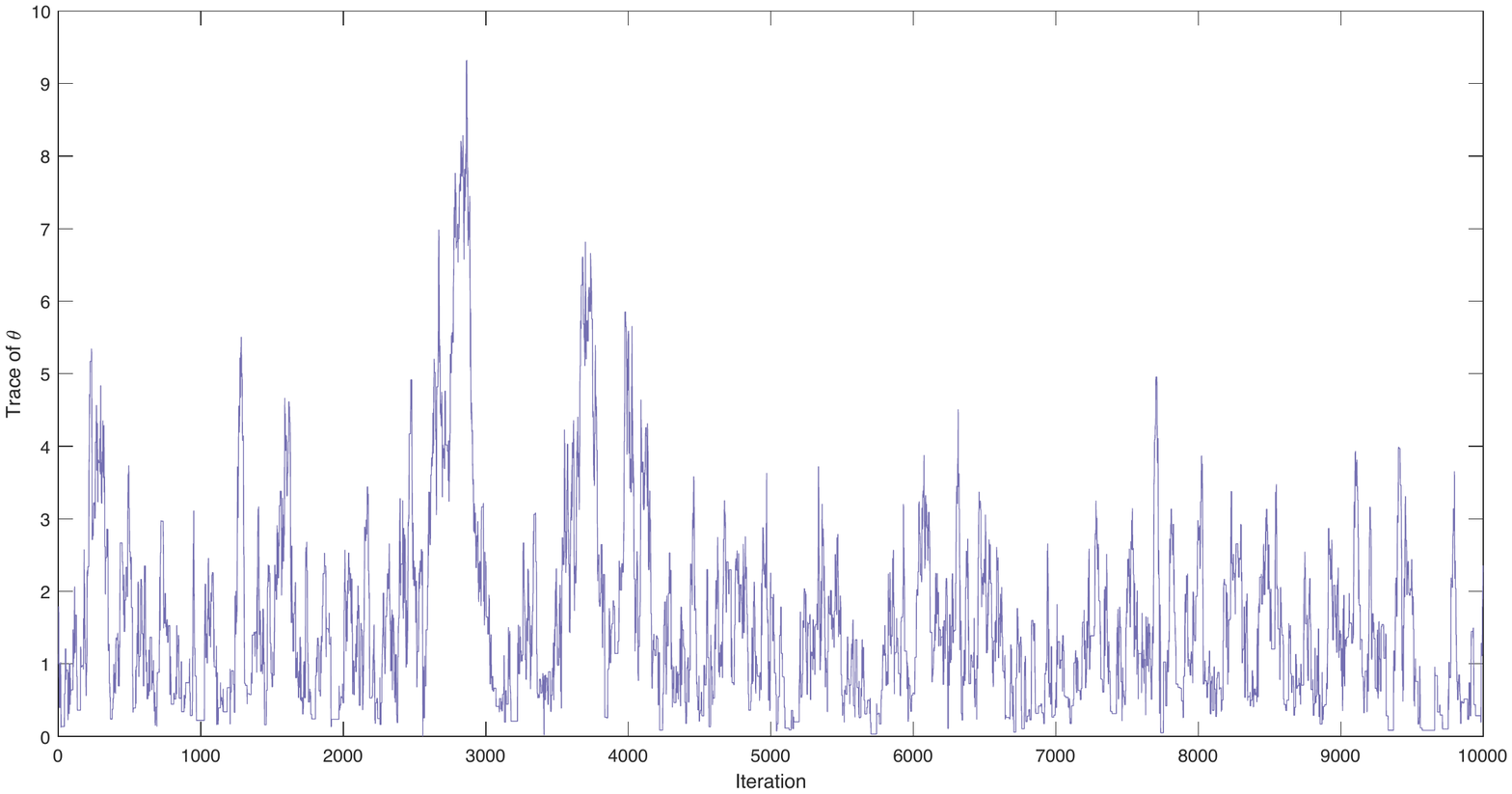}}
\subfigure[$\sigma$]{\includegraphics[width=6.5cm,height=6.5cm]{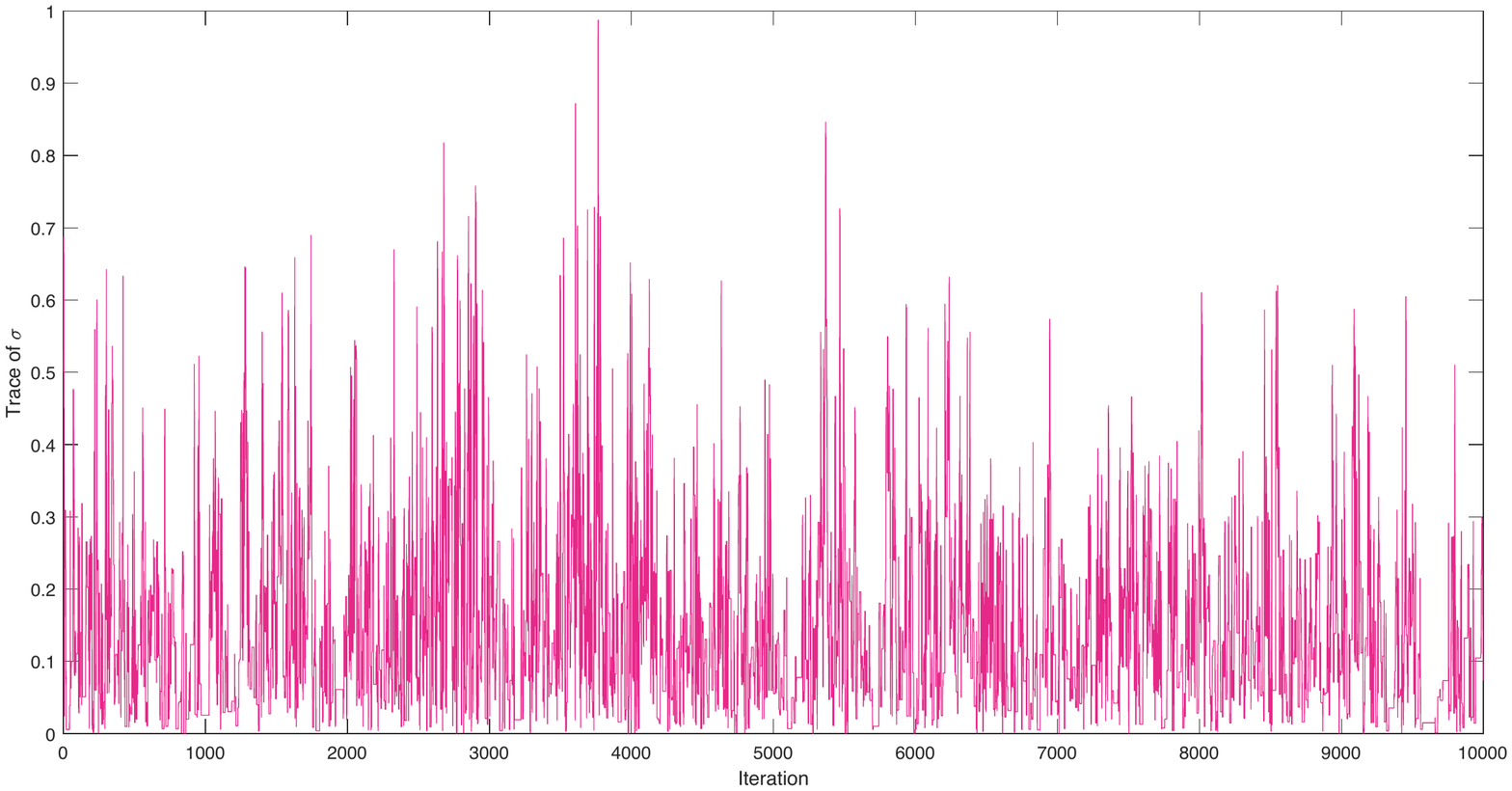}}
 \caption{Single Level Particle MCMC Results for Real Data. The simulations were at level 7.}
    \label{fig:OU2}
\end{figure}

\begin{figure}[h]
\centering
\subfigure[$\theta$]{\includegraphics[width=6.5cm,height=6.5cm]{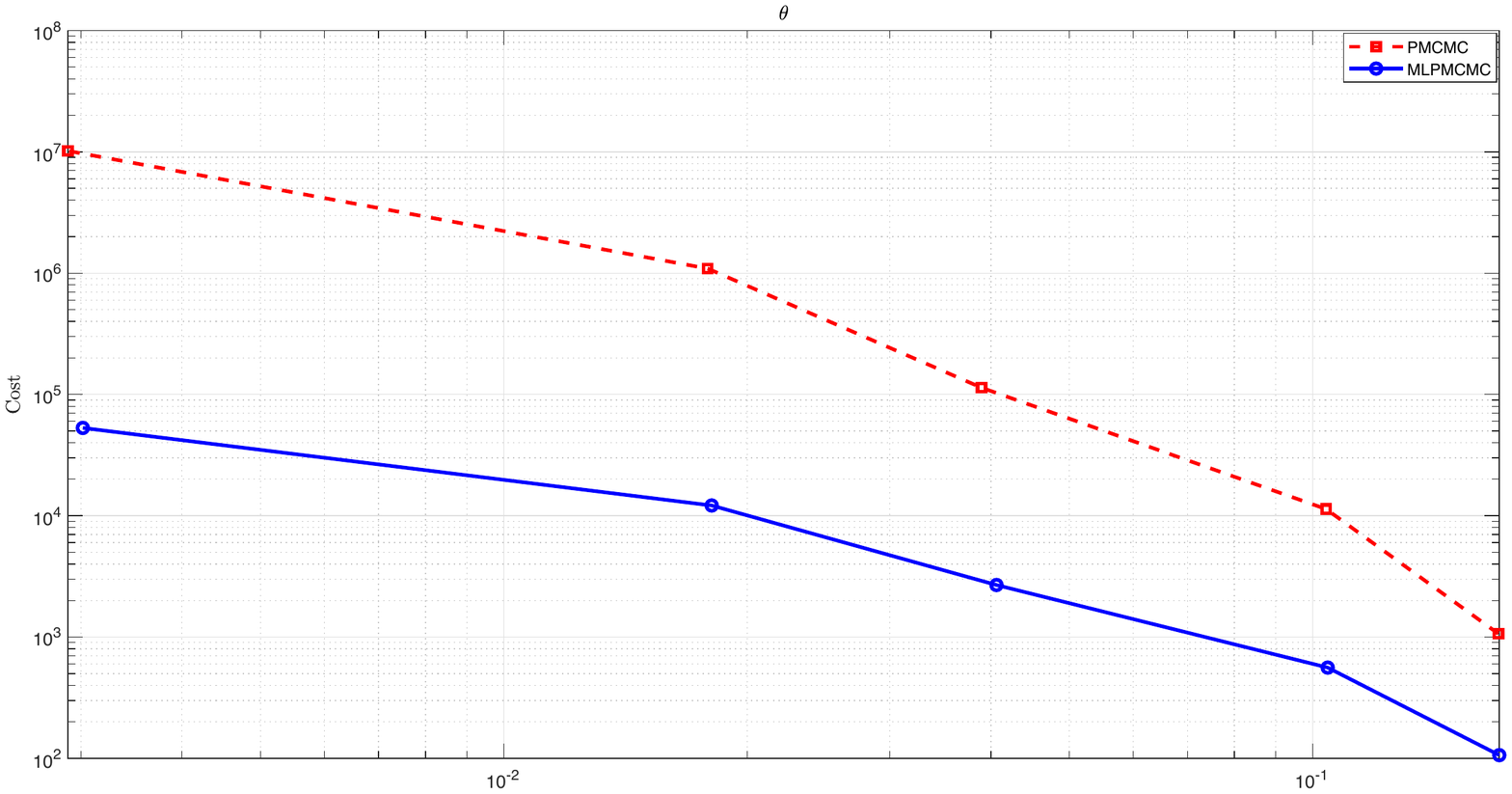}}
\subfigure[$\sigma$]{\includegraphics[width=6.5cm,height=6.5cm]{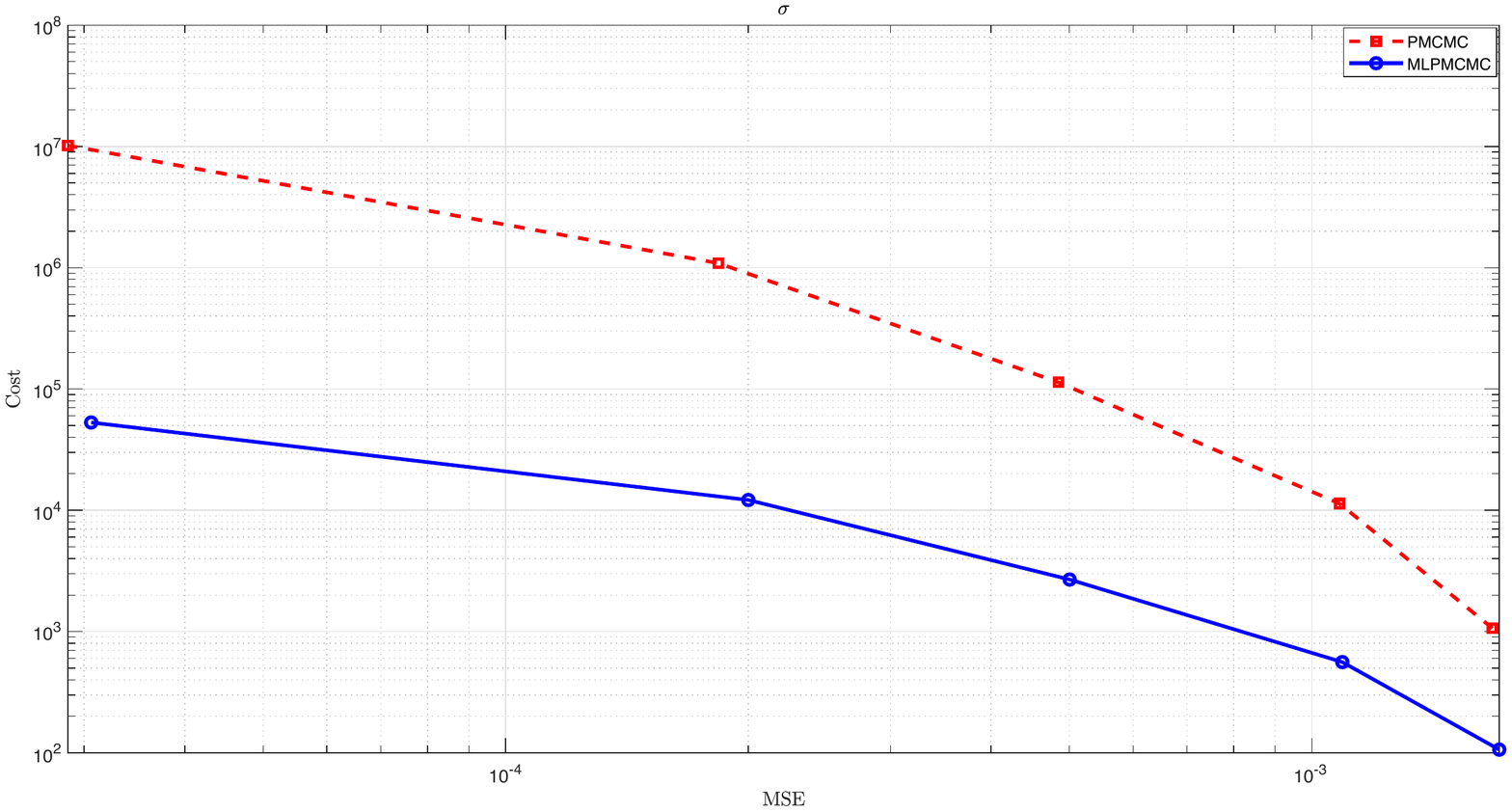}}
 \caption{Cost versus MSE Plots for Real Data. }
      \label{fig:msecost2}
\end{figure}

We repeat the experiments now for real data.  Figure \ref{fig:OU2} shows the performance of the MCMC (again single level MCMC at level 7) and is again very reasonable in terms of performance. Figure \ref{fig:msecost2} and the second row of Table \ref{tab:res}  confirm similar results for the case of simulated data, in this real data setting.

\subsubsection*{Acknowledgements}

The authors were supported by KAUST baseline funding.

\end{document}